# Critique of the Rowe 2001 Detector Efficiency Experiment

Douglas G Danforth
Greenwood Farm Technologies, LLC
Palo Alto, CA 94301
physics@greenwoodfarm.com
Version: Dec 17, 2014 (0004)

The Rowe 2001 experiment scattered photons from two prepared $^9Be^+$ ions. Measurement of the scattered photons used a single photomultiplier tube (PMT). The resultant histograms appear to be a superposition of the individual ions' distribution, but that is not correct. The PMT records the convolution of the joint probability density of the ions. There are many different joint densities which yield the same PMT histograms. Each density has a different correlation. For a fixed PMT histogram the range of those correlations can be large, e.g. -0.730 to +0.997. The reported correlations based on discriminator levels and the categories 'zero bright', 'one bright', and 'two bright' are unsupported. As such the claim that the detection efficiency loophole is closed is invalid. For this experiment, the detection loophole remains open.

It is stated in [1] that

> "In contrast to previous measurements with massive particles, this violation of Bell's inequality was obtained by use of a complete set of measurements. Moreover, the high detection efficiency of our apparatus eliminates the so-called `detection' loophole."
>
> ...
>
> The state of an ion, $|\downarrow\rangle$ or $|\uparrow\rangle$, is determined by probing the ion with circularly polarized light from a `detection' laser beam[27]. During this detection pulse, ions in the $|\downarrow\rangle$ or bright state scatter many photons, and on average about 64 of these are detected with a photomultiplier tube, while ions in the $|\uparrow\rangle$ or dark state scatter very few photons. For two ions, three cases can occur: zero ions bright, one ion bright, or two ions bright. In the one-ion-bright case it is not necessary to know which ion is bright because the Bell's measurement requires only knowledge of whether or not the ions' states are different. Figure 2 shows histograms, each with 20,000 detection measurements. The three cases are distinguished from each other with simple discriminator levels in the number of photons collected with the phototube.

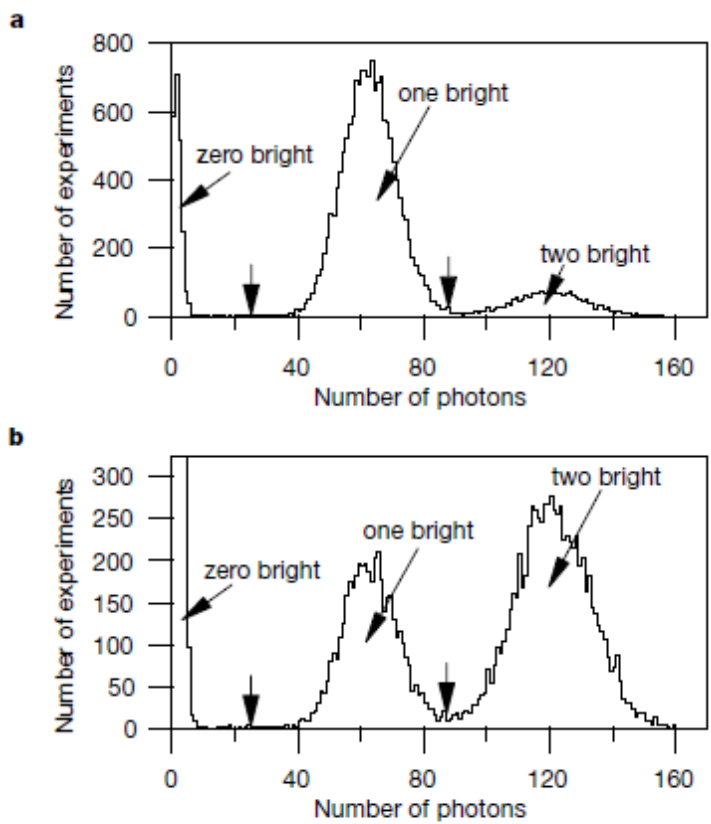

**Figure 2** Typical data histograms comprising the detection measurements of 20,000 experiments taking a total time of about 20 s. In each experiment the population in the |↓⟩ state is first coherently transferred to the $|F = 1, M_F = +1\rangle$ to make it even less likely to fluoresce upon application of the detection laser. The detection laser is turned on and the number of fluorescence photons detected by the phototube in 1 ms is recorded. The cut between the one bright and two bright cases is made so that the fractions of two equal distributions which extend past the cut points are equal. The vertical arrows indicate the location of the cut between the 0 (1) bright and 1 (2) bright peaks at 25 (86) counts. **a**, Data histogram with a negative correlation using $\phi_1 = 3\pi/8$ and $\phi_2 = 3\pi/8$. For these data $N_0 \cong 2{,}200$, $N_1 \cong 15{,}500$ and $N_2 \cong 2{,}300$. **b**, Data histogram with a positive correlation using $\phi_1 = 3\pi/8$ and $\phi_2 = -\pi/8$. For these data $N_0 \cong 7{,}700$, $N_1 \cong 4{,}400$ and $N_2 \cong 7{,}900$. The zero bright peak extends vertically to 2,551.

**Figure 0**  Histograms reported in Rowe [1]

The histograms are obtained by measurements taken from a *single* photomultiplier tube (PMT)[1]. When n photons are detected some of the photons may have come from ion 1 and some from ion 2. There are n+1 ways this can happen: (0, n), (1, n-1), (2, n-2), ..., or (n, 0) with each way having joint probability $P_{ij}$ where

$P_{ij}$ = Pr(A=i and B=j)
A = random variable of number of photons detected from ion 1
B = random variable of number of photons detected from ion 2

---

[1]Confirmed by personal communication.

Normalizing a histogram, h, over the 20,000 experiments of [1] yields a (frequency) distribution
$$f_n = h_n / \sum_{m=0}^{N} h_m = h_n/20{,}000$$
Now f arises from a *convolution* over all of the possible outcomes for a given n
$$f_n = P^*{}_n \quad \text{(convolution over joint density)}$$
where
$$P^*{}_n = \sum_{i=0}^{n} P_{i,n-i}(\phi_1, \phi_2)$$

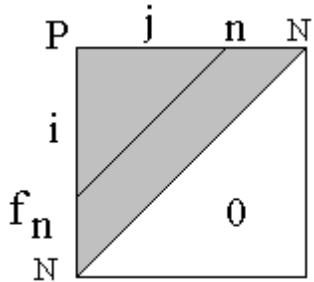

**Figure 1** nth minor diagonal of P whose sum is $f_n$

The value N is the maximum PMT count over the 20,000 experiments for a given $\phi_1$, $\phi_2$ configuration. The distribution f is not the sum of the individual distributions for ion 1 and ion 2. It is a convolution over their joint density. That understanding affects the interpretation of the discriminator levels used in [1].

The *marginals* of P are the individual ion distributions.

$$P_{i\cdot}(\phi_1, \phi_2) = \sum_{j=0}^{N} P_{ij}(\phi_1, \phi_2) \quad \text{(Ion 1 probability distribution)}$$
$$P_{\cdot j}(\phi_1, \phi_2) = \sum_{i=0}^{N} P_{ij}(\phi_1, \phi_2) \quad \text{(Ion 2 probability distribution)}$$

Experimentally all we know is f. We do not know P, only that f is related to P by a convolution. It is shown that f only weakly constrains P. There are many P that generate f.

How different can those distributions be? Write P as the sum of a symmetric and antisymmetric matrix
$$P = S + A$$
$$S^T = S \quad \text{(transpose of S)}$$
$$A^T = -A$$
The constraint that probability density P must be non negative implies
$$-S \leq A \leq S$$
One can ask for the most extreme values of A that lead to the maximal and minimal difference between the marginals $P_{i\cdot}$ and $P_{\cdot j}$ (the individual ion distributions) while holding f fixed.

If we take $A_{ij} = S_{ij}$ for $i > j$, $A_{ij} = 0$ for $i=j$, and $A_{ij} = -S_{ij}$ for $i < j$ then a maximal difference is obtained.

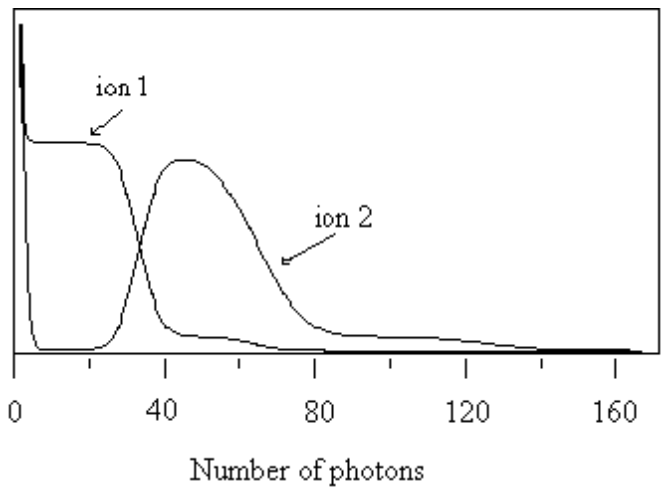

**Figure 2** Maximal marginal difference derived from Rowe Figure 2a

If we take A=0 then P is symmetric and hence the marginals are equal. One simple form for P in this case is to distribute f uniformly over the minor diagonals. P is then expressed as

$P_{ij} = f_{i+j}/(1+i+j)$   (constant minor diagonal form of P)

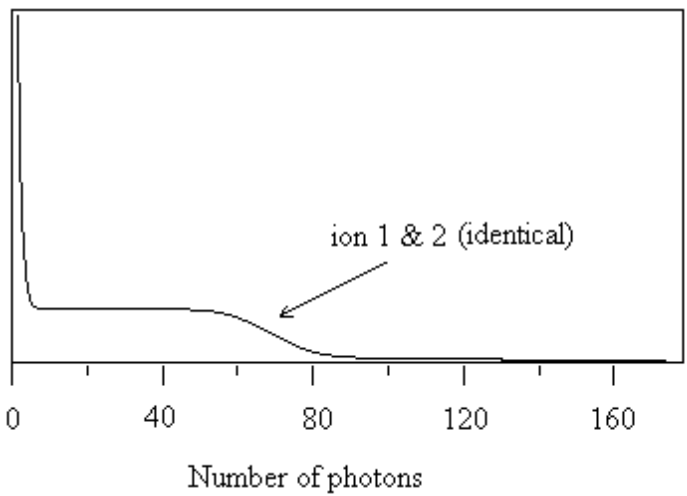

**Figure 3** A minimal marginal difference derived from Rowe Figure 2a

    It is now obvious that very little can be said concerning the individual ion distributions since the very different Figures 2 and 3 are derived from the *same* PMT histogram of Rowe Figure 2a. Just because there are hills and valleys in the histogram does not mean that those hills and valleys are present in the marginal distributions.
    Three situations are considered that bring forth the range of uncertainty in the ions' correlations when only a single photomultiplier tube is used to record photons from both ions.

(Edge) negatively correlationed:
$P_{ij} = 1/2 f_{i+j}(\delta_{i,0} + \delta_{j,0})$

(Random) M is a random non negative matrix whose convolution is not zero:

$$P_{ij} = f_{i+j} M_{ij} / M^*_{i+j}$$

(Diagonal) positively correlationed:
$$P_{ij} = \delta_{i,j} \quad i+j \text{ even,}$$
$$P_{ij} = 1/2\, f_{i+j}(\delta_{i,j-1} + \delta_{i,j+1}) \quad i+j \text{ odd.}$$

The histograms of Rowe Figure 2a and 2b are used for the 3 forms and their correlations shown in table 1

Table 1 Possible correlations of the histograms of [1]

|  | Fig 2a | Fig 2b |
|---|---|---|
| Edge | -0.7206 | -0.3052 |
| Random | -0.3534 | 0.2286 |
| Diagonal | 0.9997 | 0.9999 |

The uncertainty in correlation of $r(\varphi_1, \varphi_2)$ is at least as large as
$$-0.7206 \leq r(3\pi/8, 3\pi/8) \leq 0.9997 \quad \text{Fig 2a}$$
$$-0.3052 \leq r(3\pi/8, -\pi/8) \leq 0.9999 \quad \text{Fig 2b}$$

Notice that the correlations were derived without reference to any discriminator levels and used possible joint densities that preserved the histograms. The question now becomes whether the discriminator levers bear any connection to the actual ion correlations.

Two discriminator levels, $t_1$ and $t_2$, are used in [1] to partition the histograms into three classes: 'zero bright', 'one bright' and 'two bright'. Those labels are problematic as can be seen in Figure 4 which shows two different distributions that *maintain* the histograms but which have very different interpretations.

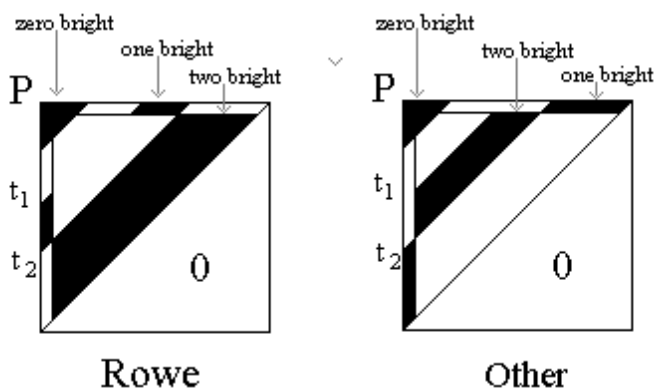

Figure 4  Two interpretations of discriminator levels

The 'Rowe' interpretation considers that all of the probability mass is consentrated between $t_1$ and $t_2$ along the edge of the density and for counts greater than $t_2$ the probability mass is consentrated at the interior of the density. The 'zero bright', 'one bright' and 'two bright' regions are so marked. The 'Rowe' interetation asserts that for modest counts the ions act disjoint from each other, one is on while the other is off.  For high counts the 'Rowe' interpretation asserts that both ions are on.

The 'Other' interpretation concentrates the mass between $t_1$ and $t_2$ at the interior while for counts greater than $t_2$ the mass is concentrated at the edges.  The labels for the regions are so marked.  The 'Other' interpretation asserts that for modest counts both ions are on and that for high counts the ions alternate who is on.  The 'Other' interpretation flips the labels of one and two from the 'Rowe' interpretation.  This has a profound effect on the correlations based on those categories.

Blends between 'Rowe' and 'Other' are possible and lead to wide variations in the correlations calculated.  In the blended case the categories 'one' and 'two' bright become muddled.  The classification of regions is open to more than one interpretation.

The ambiguity in the correlations can be removed by redesigning and rerunning the experiment with two photo detectors each accepting photons from only one of the ions.  The joint probability density P is then directly measureable.  No thresholds are necessary.  The CHSH inequality is applicable to integer measurements, 0, ..., N if one simply divides all counts by N (the maximum count over all runs).  The data is then bounded by 1 and the CHSH derivation can be used [2].

### Further Analysis
Using an artificial genetics search algorithm and the assumption that the ions are acting *independently* it is possible to *deconvolve* the histograms of Figure 0 and obtain the histogram of each ion separately.  Figure 5 shows the deconvolution of Rowe Figure 2a.  Each ion is shown in a different color (blue and green). The horizontal axis is the intensity of the scattered photons.  The vertical axis is the probability of obtaining the specified intensity.

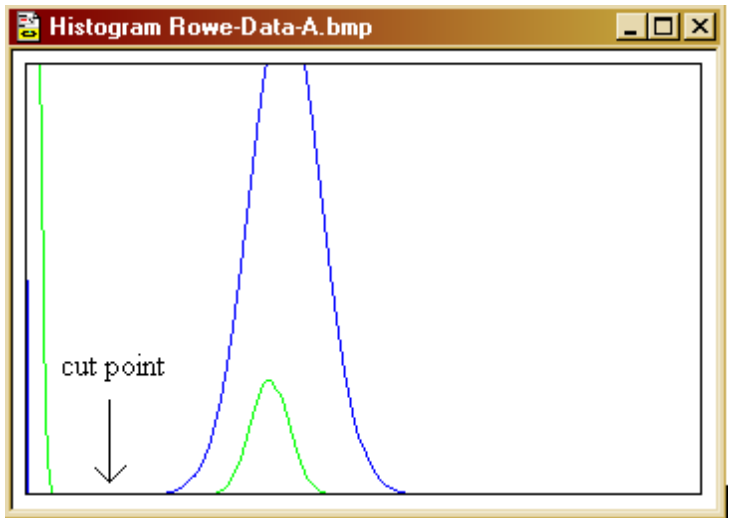

**Figure 5** Deconvolved ion 1 and 2 histograms for Rowe Figure 2a.

Looking at the green curve one can see there are two distinct parts, one with low intensity (dark) and one with high intensity (bright).  There is a 'cut' point between the dark and bright that can completely disambiguate the two states of that ion (and similarly for the blue curve).  In statistics that process is called discriminant analysis [3].  The probability of correct classification (PCC) for the cases in Figure 5 is essentially 1.   When one convolves those two ions one regenerates the Rowe Figure 2a.  The red curve in Figure 6 is the regenerated data and the black curve is the original data.

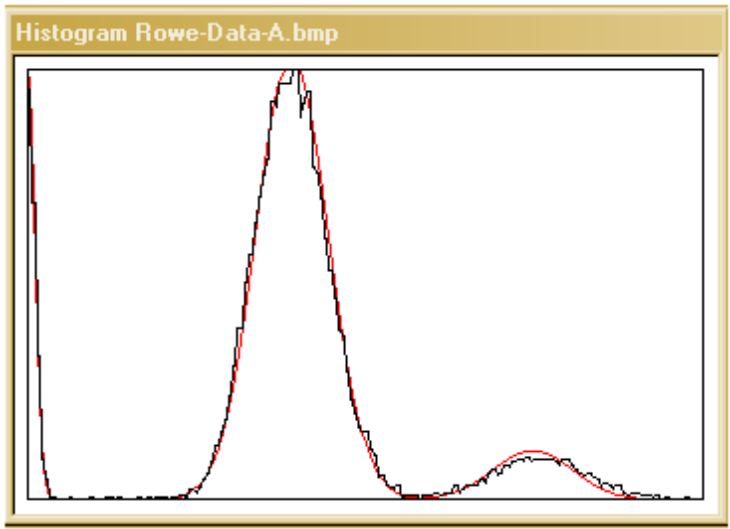
**Figure 6** Convolved ion distributions regenerating original data of Rowe Figure 2a.

Similarly for the Rowe Figure 2b histogram we obtain the following individual ion distributions (blue and green). Please note that "the zero bright peak extends vertically to 2,551". That extension has been included in the Rowe Figure 2b data which has the effect of reducing the size of the individual peaks relative to the zero bright peak.

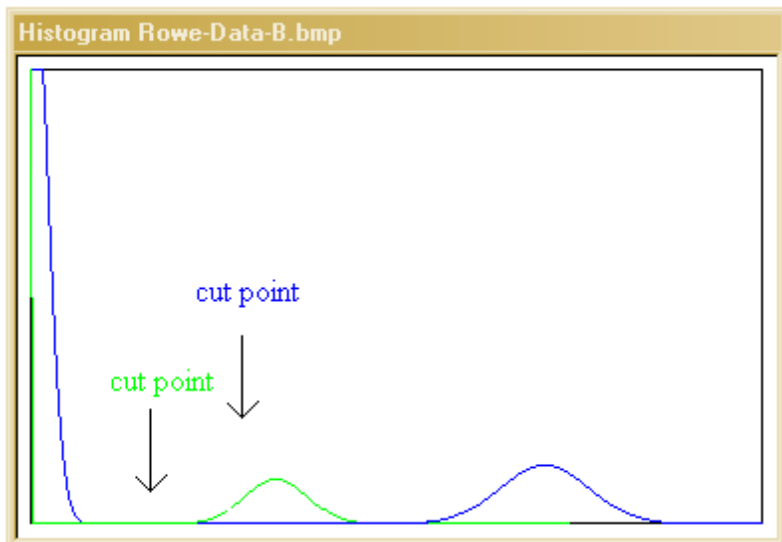
**Figure 7** Deconvolved ion 1 and 2 histograms for Rowe Figure 2b.

Again using those individual ion distribution one can reproduce the original data as shown in Figure 8 where the reproduction is in red and the original is in black.

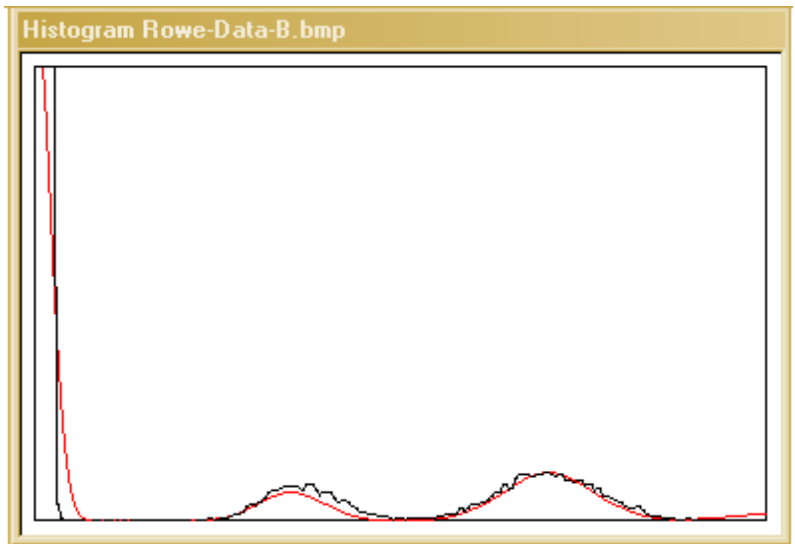
**Figure 8** Convolved ion distributions regenerating original data of Rowe Figure 2b.

**Summary**
The cut points clearly discriminate 'dark' and 'bright' states. But that is insufficient. *Independent* ions with *zero* correlation are capable of reproducing the Rowe histograms. This shows that their threshold calculations are invalid.

The data collected from a single photomultiplier for two beryllium 9 ions does not have sufficient information to unambiguously calculate the ions' statistics. As such the conclusion drawn by the authors is unwarrented. The Rowe 2001 experiment does not close the detection efficiency loophole.

______________________________________________________________________________